\documentclass[preprint,showpacs,12pt,floatfix,aps]{revtex4}
 
\usepackage{graphicx}
\usepackage{latexsym}
\usepackage{amsmath}
\input psfig.sty

\newcommand{\bq}{\begin{equation}}
\newcommand{\eq}{\end{equation}}
\newcommand{\bqn}{\begin{eqnarray}}
\newcommand{\eqn}{\end{eqnarray}}
\newcommand{\nb}{\nonumber}
\newcommand{\lb}{\label}
\newcommand{\rr}{\bf r}

\begin{document}
\title{Gravastars and Black Holes of Anisotropic Dark Energy}
\author{R. Chan $^{1}$}
\email{chan@on.br}
\author{M.F.A. da Silva $^{2}$}
\email{mfasnic@gmail.com}
\author{P. Rocha $^{2,3}$}
\email{pedrosennarocha@gmail.com}
\affiliation{\small $^{1}$ Coordena\c{c}\~ao de Astronomia e Astrof\'{\i}sica, 
Observat\'orio Nacional, Rua General Jos\'e Cristino, 77, S\~ao Crist\'ov\~ao  
20921-400, Rio de Janeiro, RJ, Brazil\\
$^{2}$ Departamento de F\'{\i}sica Te\'orica, Instituto de F\'{\i}sica, 
Universidade do Estado do Rio de Janeiro, Rua S\~ao Francisco Xavier 524, 
Maracan\~a 20550-900, Rio de Janeiro - RJ, Brasil\\
$^3$ Ger\^encia de Tecnologia da Informa\c c\~ao, ACERP, TV Brasil,
R\'adios Nacional e MEC,
Rua da Rela\c c\~ao 18,  Lapa, CEP 20231-110, Rio de Janeiro, RJ, Brazil}
 
\date{\today}

\begin{abstract}
Dynamical models of prototype gravastars made of anisotropic dark energy are constructed, in 
which an infinitely thin spherical shell of a perfect fluid with the equation of 
state $p = (1-\gamma)\sigma$ divides the whole spacetime into two regions, the 
internal region filled with a dark energy fluid, and the external Schwarzschild region. 
The  models represent "bounded excursion"  
stable gravastars, where  the thin shell is oscillating between two finite radii, 
while in other cases they collapse until the formation of black holes. 
Here we show, for the first time in the literature, a model of gravastar and
formation of black hole with 
both interior and thin shell constituted exclusively
of dark energy. 
Besides, the sign of the parameter of
anisotropy ($p_t - p_r$) seems to be relevant to the gravastar formation. 
The formation is favored when the
tangential pressure is greater than the radial pressure, at least in the
neighborhood of the isotropic case ($\omega=-1$).
  
\end{abstract}


\maketitle

\section{Introduction}

Gravastar was proposed as an alternative to black holes.
The pioneer of Mazur and Mottola (MM) \cite{MM01},
consists of five layers: an internal core
$0 < r < r_1$ , described by the de Sitter universe, an intermediate thin layer of stiff fluid
$r_1 < r < r_2$ , an external region $r > r_2$ , described by the Schwarzschild solution, and two 
infinitely thin shells, appearing, respectively, on the hypersurfaces $r = r_1$ and
$r = r_2$. The intermediate layer is constructed in such way that $r_1$ is inner than the de Sitter horizon, while $r_2$ is outer than the Schwarzschild horizon, eliminating the apparent horizon. Configurations with a de Sitter interior have long history which we can find, for example, in the work of Dymnikova and Galaktionov \cite{irina}.  
After this work, Visser and Wiltshire \cite{VW04} pointed out that there are 
two different types of stable gravastars which are stable gravastars and 
"bounded excursion" gravastars. In the spherically symmetric case, the motion 
of the surface of the gravastar can be written in the form \cite{VW04},
\bq
\lb{1.4}
\frac{1}{2}\dot{R}^{2} + V(R) = 0,
\eq
where $R$ denotes the radius of the star, and $\dot{R} \equiv dR/d\tau$, with 
$\tau$ being the proper time of the surface. Depending on the properties of 
the potential $V(R)$, the two kinds of gravastars are defined as follows. 

{\bf Stable gravastars}: In this case,  there must exist a radius $a_{0}$ such that
\bq
\lb{1.5}
V\left(R_{0}\right) = 0, \;\;\; V'\left(R_{0}\right) = 0, \;\;\;
V''\left(R_{0}\right) > 0,
\eq
where a prime denotes the ordinary differentiation with respect to the indicated argument.
If and only if there exists such a radius $R_{0}$ for which the above conditions are satisfied,
the model is said to be stable. 
Among other things, VW found that there are many equations of state for which the gravastar
configurations are stable, while others are not \cite{VW04}. Carter studied  the same
problem and found new equations of state for which the gravastars are stable \cite{Carter05}, 
while De Benedictis {\em et al} \cite{DeB06} and Chirenti and Rezzolla \cite{CR07} 
investigated the stability of the original model
of  Mazur and  Mottola against axial-perturbations, and found that gravastars are stable to
these perturbations too. Chirenti and Rezzolla also showed that their quasi-normal modes 
differ from those of black holes with the same mass, and thus can be used to discern a gravastar 
from a black hole. 

{\bf "Bounded excursion" gravastars}: As VW noticed, there is a less stringent notion of 
stability, the so-called "bounded excursion" models, in which there exist two radii $a_{1}$ 
and $a_{2}$ such that
\bq
\lb{1.6}
V\left(R_{1}\right) = 0, \;\;\; V'\left(R_{1}\right) \le 0, \;\;\;
V\left(R_{2}\right) = 0, \;\;\; V'\left(R_{2}\right) \ge 0,
\eq
with $V(R) < 0$ for $R \in \left(R_{1}, R_{2}\right)$, where $R_{2} > R_{1}$. 

Lately, we studied both types of gravastars \cite{JCAP}-\cite{JCAP4}, and found that, 
such configurations can indeed be constructed, although   the region for the formation 
of them is very small in comparison to that of black holes.

Based on the discussions about the gravastar picture some authors have proposed
alternative models \cite{Chan}. Among them, we can find a Chaplygin dark star \cite{Paramos},
a gravastar supported by non-linear electrodynamics \cite{Lobo07},
a gravastar with continuous anisotropic pressure \cite{CattoenVisser05}
and recently, Dzhunushaliev et al. worked on spherically symmetric configurations
of a phantom scalar field and they found something like a gravastar but it was unstable\cite{singleton}.
In addition, Lobo \cite{Lobo} studied two models for a dark energy fluid. One of them
describes a homogeneous energy density and the other one describes  an
ad-hoc monotonically decreasing energy density, although both of them are with anisotropic
pressure.  In order to match an exterior Schwarzschild spacetime he  
introduced a thin shell between the interior and the exterior spacetimes.

Since in the study of the evolution of gravastar there is a possibility
of black hole formation, we can find some works considering the hypothesis
of dark energy black hole. In particular, Debnath and Chakraborty \cite{Debnath}
studied the collapse of a spherical cloud, consisting of both dark matter and 
dark energy, in a form of modified Chaplygin gas.  They have found when the
collapsing fluid is only formed by dark energy, the final stage is always a
black hole.  On the other hand, Cai and Wang \cite{Cai} studying the collapse
of a spherically symmetric star, made of a perfect fluid, have found that if
the fluid would correspond to a dark energy, black hole would never be formed.

Here we are interested in the study of a gravastar model whose
interior consists of an anisotropic dark energy fluid \cite{Lobo} which
admits the isotropy as a particular case, 
and analyze the effect of the
anisotropy of the pressure in the evolution of gravastars.
We shall first construct three-layer dynamical models,  and then show that both types of
gravastars and black holes exist for various situations. 
We have also shown a model of gravastar and
formation of black hole with
both interior and thin shell constituted exclusively
of dark energy.
The rest of 
the paper is  organized as follows: In Sec. II we present the metrics of the 
interior and exterior spacetimes, and write down the motion of the thin shell
in the form of  equation (\ref{1.4}).  In Sec. III we show the definitions of dark and 
phantom energy, for the
interior fluid and for the shell.  From Sec. IV  we discuss the formation of
black holes and gravastars for anisotropic and isotropic fluids of 
different kinds of energy.
Finally, in Sec. V we present our conclusions.

\section{ Dynamical Three-layer Prototype Gravastars}

The interior fluid is made of an anisotropic dark energy fluid with a metric
given by the first Lobo's model \cite{Lobo}
\bq
ds^2_{i}=-f_1 dt^2 + f_2 dr^2 + r^2 d\Omega^2,
\lb{ds2-}
\eq
where  $d\Omega^2 \equiv d\theta^2 + \sin^2(\theta)d\phi^2$, and 
\bqn
f_1 &=& (1-2 a r^2)^{-\frac{1+3\omega}{2}},\nb\\
f_2 &=& \frac{1}{1 -2 a r^2},
\eqn  
where $\omega$ is a constant, and its physical meaning can be seen from the
following equation (\ref{prpt}). Since the mass is given by 
$\bar m(r)=4\pi\rho_0 r^3/3$ and $a=4\pi\rho_0/3$ then we have that $a > 0$,
where $\rho_0$ is the homogeneous energy density.
Note that there is a horizon at $r_h=1/\sqrt{2a}$, thus the radial coordinate
must obey $r < r_h$. 
The corresponding energy density $\rho$, radial  and tangential  pressures $p_r$ and 
$p_t$ are given, respectively, by
\bqn
\rho&=&\rho_0=constant, \nb \\
p_r&=&\omega \rho_0, \nb \\
p_t&=&\omega \rho_0 \left[ 1 + \frac{4\pi}{6} \frac{(1+\omega)(1+3\omega)\rho_0 r^
2}
{\omega \left(1- \frac{8\pi \rho_0}{3}r^2 \right)} \right], \\
\lb{prpt}
\end{eqnarray}
when $\omega=-1$ and $\omega=-1/3$ we obtain an interior isotropic pressure fluid.

The exterior spacetime is given by the Schwarzschild metric
\bq
ds^2_{e}= - f dv^2 + f^{-1} d{\rr}^2 + {\rr}^2 d\Omega^2,
\lb{ds2+}
\eq
where $f=1 - {2m}/{\rr}$.
The metric of the hypersurface  on the shell is given by
\bq
ds^2_{\Sigma}= -d\tau^2 + R^2(\tau) d\Omega^2,
\lb{ds2Sigma}
\eq
where $\tau$ is the proper time.

Since $ds^2_{i} = ds^2_{e} = ds^2_{\Sigma}$, we find that  $r_{\Sigma}={\rr}_{\Sigma}=R$,
and  
\bqn
\lb{dott2}
f_1\dot t^2 - f_2 \dot R^2 &=&  1,\\
\lb{dotv2}
f\dot v^2 - \frac{\dot R^2}{f} &=& 1,
\eqn
where the dot denotes the ordinary differentiation with respect to the proper time.
On the other hand,  the interior and exterior normal vectors to the thin shell are given by
\bqn
\lb{nalpha-}
n^{i}_{\alpha} &=& (-\dot R, \dot t, 0 , 0 ),\nb\\
n^{e}_{\alpha} &=& (-\dot R, \dot v, 0 , 0 ).
\eqn
Then, the interior and exterior extrinsic curvatures are given by
\bqn
K^{i}_{\tau\tau} &=& -(1-2 a R^2)^{-(3\omega+1)/2} \left\{ \left[ 6 (1-2 a 
R^2)^{(3\omega+1)/2} \dot R^2 \omega+6 a R^2 \dot t^2 \omega+2 a 
R^2 \dot t^2-3 \dot t^2 \omega-\dot t^2 \right]  \right. \nb \\
& &\left. \times a R \dot t-(1-2 a R^2)^ {(3 
\omega+1)/2} (-1+2 a R^2) (\dot R \ddot t-\ddot R \dot t)\right\} (-1+2 a R^2)
^{-1}, \\
\lb{Ktautau-}
K^{i}_{\theta\theta} &=&
\dot t(1-2a R^2) R,\\
\lb{Kthetatheta-}
K^{i}_{\phi\phi} &=& K^{i}_{\theta\theta}\sin^2(\theta),\\
\lb{Kphiphi-}
K^{e}_{\tau\tau} &=& \dot v(4 m^2 \dot v^2-4 m R \dot v^2-3 R^2 \dot R^2+
R^2 \dot v^2) (2 m-R)^{-1} m R^{-3}+\dot R \ddot v- \ddot R \dot v,\\
\lb{Ktautau+}
K^{e}_{\theta\theta} &=& -\dot v (2 m-R),\\
\lb{Kthetatheta+}
K^{e}_{\phi\phi} &=& K^{e}_{\theta\theta}\sin^2(\theta).
\lb{Kphiphi+}
\eqn
Since \cite{Lake}
\bq
[K_{\theta\theta}]= K^{e}_{\theta\theta}-K^{i}_{\theta\theta} = - M,
\lb{M}
\eq
where $M$ is the mass of the shell, we find that
\bq
M=\dot v (2 m-R)+\dot t(1-2a R^2) R.
\lb{M1}
\eq
Then, substituting equations (\ref{dott2}) and (\ref{dotv2}) into (\ref{M1}) 
we get
\bq
M=-R\left(1-\frac{2m}{R} + \dot R^2 \right)^{1/2} + 
R\frac{ \left( 1 -2a R^2 + \dot R^2  \right)^{1/2}}
{(1-2a R^2)^{-(3\omega+2)/2}}.
\lb{M2}
\eq
In order to keep the ideas of MM as much as possible, we consider the thin 
shell as consisting
of a fluid with the equation of state, $p=(1-\gamma)\sigma$, where $\sigma$ and $p$ denote, 
respectively, the surface energy density and pressure of the shell and $\gamma$ is a constant. 
Then, the equation of motion of the shell is given by \cite{Lake}
\bq
\dot M + 8\pi R \dot R p = 4 \pi R^2 [T_{\alpha\beta}u^{\alpha}n^{\beta}]=
\pi R^2 \left(T^e_{\alpha\beta}u_e^{\alpha}n_e^{\beta}-T^i_{\alpha\beta}u_i^{\alpha}n_i^{\beta} \right),
\lb{dotM}
\eq
where $u^{\alpha}$ is the four-velocity.  Since the interior fluid is made
of an anisotropic fluid and the exterior is vacuum, we get
\bq
\dot M + 8\pi R \dot R (1-\gamma)\sigma = 0.
\lb{dotM1}
\eq
Recall that $\sigma = M/(4\pi R^2)$, we find that equation (\ref{dotM1}) has the solution
\bq
M=k R^{2(\gamma-1)},
\lb{Mk}
\eq
where $k$ is an integration constant. Substituting equation (\ref{Mk}) into equation (\ref{M2}),
and rescaling $m, \; b$ and $R$ as,
\bqn
m &\rightarrow& mk^{-\frac{1}{2\gamma-3}},\nb\\
a &\rightarrow& a k^{\frac{2}{2\gamma-3}},\nb\\
R &\rightarrow& Rk^{-\frac{1}{2\gamma-3}},
\eqn
we find that it can be written in the form of  equation (\ref{1.4}) 
\bqn
V(R,m,\omega,a,\gamma)&=& 
\frac{1}{2 R^2 (b^2-1)}\left\{ R^2 b^2-R^2+2 a R^4+b^2 R^{4 \gamma-4}+b^2 R^{2\gamma-2} \left[-2 b^2 R^{2\gamma-2}- \right. \right. \nb \\
& &\left. \left. 2 \sqrt{2 b^2 m R-2 b^2 a R^4-2 m R+2 a R^4+b^2 R^{4\gamma-4}}\right](b^2-1)^{-1}-2 b^2 m R \right\}, \nb \\
\lb{VR}
\eqn
where
\bq
\lb{b1}
b \equiv (1-2a R^2)^{-(1+3\omega/2)}.
\eq
Clearly, for any given constants $m$, $\omega$, $a$ and $\gamma$, equation (\ref{VR}) uniquely 
determines the collapse of the prototype  gravastar. Depending on the initial value $R_{0}$,  
the collapse can form either a black hole,  a gravastar or a spacetime filled with
anisotropic dark energy fluid. In the last case, the thin shell
first collapses to a finite non-zero minimal radius and then expands to infinity.  To  guarantee
that initially the spacetime does not have any kind of horizons,  cosmological or event,
we must restrict $R_{0}$ to the range,
\bq
\lb{2.2b}
2m < R_{0} < 1/\sqrt{2a},
\eq
where $R_0$ is the initial collapse radius. 

Solving $V(R,m,\omega,a,\gamma)=0$ with respect to $m$, we can find the
critical mass given by 
\bqn
m_c&=& \frac{1}{b^2 R}\left[\frac{1}{2} R^2 b^2-\frac{1}{2} b^2 R^{4\gamma-4}+a R^4-\frac{1}{2} R^2-\sqrt{-2 a b^2 R^{4\gamma}+ b^2 R^{4\gamma-2}}\right].
\eqn

Since the potential  (\ref{VR}) is so complicated, it is too difficult to study it
analytically. Instead, in the following we shall study it numerically. 

\section{Classifications of Matter, Dark Energy, and Phantom Energy  for Anisotropic Fluids}

Recently \cite{Chan08}, an explicit classification of matter, dark and phantom energy 
for an anisotropic fluid was given  in terms of the energy conditions. Such a classification
is necessary for systems where anisotropy is important, and  the pressure components 
may play very important roles and  can have quite different contributions.
In this paper, we will consider this classification to study the collapse of the
dynamical prototype gravastars, constructed in the last section. 
In particular, we define dark
energy  as a fluid which violates the strong energy
condition
(SEC).  From the Raychaudhuri equation, we can see that such defined
dark energy always exerts  divergent forces on  time-like or null geodesics.
On the other hand,  we define phantom energy as a fluid  that violates at least
one of the null energy conditions (NEC's). We shall further distinguish phantom
energy that satisfies the SEC
from that which does not satisfy the SEC. We call
the former attractive 
phantom energy, and the latter
 repulsive phantom energy.
Such a classification is summarized in Table I.

For the sake of completeness, in Table II we apply it to the matter field
located on the thin shell, while in Table III we combine all the results of Tables I 
and II, and present all the possibilities found.   

\begin{table}
\caption{\label{tab:table1} This table summarizes the  classification of
the interior matter field, based on the energy conditions \cite{HE73}, where 
we assume that $\rho \ge 0$.}
\begin{ruledtabular}
\begin{tabular}{cccc}
Matter & Condition 1 & Condition 2  & Condition 3 \\
\hline
Normal Matter           & $\rho+p_r+2p_t\ge 0$ & $\rho+p_r\ge 0$ & $\rho+p_t\ge 0$ \\
\hline
Dark Energy               & $\rho+p_r+2p_t <  0$ & $\rho+p_r\ge 0$ & $\rho+p_t\ge 0$ \\
\hline
                          &                      & $\rho+p_r <  0$ & $\rho+p_t\ge 0$ \\
Repulsive Phantom Energy  & $\rho+p_r+2p_t <  0$ & $\rho+p_r\ge 0$ & $\rho+p_t <  0$ \\
                          &                      & $\rho+p_r <  0$ & $\rho+p_t <  0$ \\
\hline
                          &                      & $\rho+p_r <  0$ & $\rho+p_t\ge 0$ \\
Attractive Phantom Energy & $\rho+p_r+2p_t\ge 0$ & $\rho+p_r\ge 0$ & $\rho+p_t <  0$ \\
                          &                      & $\rho+p_r <  0$ & $\rho+p_t <  0$ \\
\end{tabular}
\end{ruledtabular}
\end{table}

\begin{table}
\caption{\label{tab:table2} This table summarizes the  classification of matter
on the thin shell, based on the energy conditions \cite{HE73}. The last column indicates
the particular values of the parameter $\gamma$, where we assume that $\rho \ge 0$.}
\begin{ruledtabular}
\begin{tabular}{cccc}
Matter & Condition 1 & Condition 2  & $\gamma$ \\
\hline
Normal Matter           & $\sigma+2p\ge 0$ & $\sigma+p\ge 0$ & -1 or 0  \\
Dark Energy               & $\sigma+2p <  0$ & $\sigma+p\ge 0$ &  7/4 \\
Repulsive Phantom Energy  & $\sigma+2p <  0$ & $\sigma+p <  0$ &   3  \\
\end{tabular}
\end{ruledtabular}
\end{table}

In order to consider the equations (\ref{ds2-}) and (\ref{prpt}) for describing dark energy
stars we must analyze carefully the ranges of the parameter $\omega$ that in
fact furnish the expected fluids.  It can be shown that the
condition $\rho+p_r>0$ is violated for $\omega<-1$ and fulfilled for $\omega>-1$, 
for any values of $R$ and $a$.
The conditions $\rho+p_t>0$ and $\rho+p_r+2p_t>0$ are satisfied for $\omega<-1$
and $-1/3<\omega<0$, for any values of $R$ and $a$.  
For the other intervals of
$\omega$  the
 energy conditions depend
on very complicated relations of $R$ and $a$.  See \cite{Chan08}. 
This provides an explicit example, in which the definition of dark energy must be
dealed with great care.  Another case was provided in a previous work \cite{Chan08}.  
In a recent paper we have considered several values of $\omega$ in the intervals $-1<\omega<-1/3$ and 
$\omega>0$ for an isotropic interior model and we could not found any 
case where the interior dark energy exists, in contrast to our results presented
in this work.

In order to fulfill the energy condition $\sigma+2p\ge0$ of the shell
and assuming that
$p=(1-\gamma)\sigma$ we must have $\gamma \le 3/2$. On the other hand, in order
to satisfy the condition $\sigma+p\ge 0$, we obtain $\gamma \le 2$.
Hereinafter, we will use only some particular values of the parameter
$\gamma$ which are analyzed in this work. See Table II.

In the next sections we will discuss the different types of physical systems 
that we can find in the study of the potential $V(R,m,\omega,a,\gamma)$.

\section{Possible Configurations}

Here we can find many types of systems, depending on the combination of the 
constitution matter of the shell and core.  Among them, there are formation of
black holes, stable gravastars and dispersion of the shell, as it has already 
shown in our previous works \cite{JCAP}-\cite{JCAP4} and listed in the table III. 

As can be seen in the figures \ref{fig2}, \ref{fig3}, \ref{fig4} and \ref{fig5},
the formation of the gravastar appears as the unique possibility, for a given 
choice of the physical parameters. We can see that $V(R) = 0$ now can have  one,
two  or three real roots, depending on the mass of the shell. For $m>m_c$ we 
have, say, $R_{i}$, where $R_{i+1} > R_{i}$. If we choose $R_{0} > R_{3}$ 
(for $m=m_c$ we have $R_2=R_3$), then the star will not be allowed in this 
region because the potential is greater than the zero.  However,
if we choose $R_{1} < R_{0} < R_{2}$, the collapse will bounce back and forth 
between $R = R_{1}$ and $R = R_{2}$. Such a  possibility is shown in these 
figures. This is exactly the so-called "bounded excursion" model mentioned in 
\cite{VW04}, and studied in some details in \cite{JCAP}-\cite{JCAP4}.  Of 
course, in a realistic situation, the star will emit both gravitational waves 
and particles, and the potential will be self-adjusted to produce a minimum at 
$R = R_{static}$ where 
$V\left(R=R_{static}\right) = 0 = V'\left(R=R_{static}\right)$ 
whereby a gravastar is finally formed \cite{VW04,JCAP,JCAP1,JCAP2}.

Moreover, the model considered now, allows us to investigate if 
the anisotropy has a relevant role in the gravastar formation. The graphics 
\ref{fig2}, \ref{fig3}, \ref{fig4} and \ref{fig5} also
suggest that gravastar is formed for isotropic as well as anisotropic fluid 
cores. However, it is interesting to note that the sign of the parameter of 
anisotropy ($p_t - p_r$) seems to be relevant to this formation. Those figures 
indicate, in a first view, that gravastar formation is favored when the 
tangential pressure is greater than the radial pressure, at least in the 
neighborhood of the isotropic case ($\omega=-1$).

Another two completely new configurations appear from the potential studied 
here. They are represented by figures \ref{fig7} and \ref{fig8} and 
corresponding to a gravastar and a black hole, respectively, formed uniquely by 
non standard energy. The former one corresponds to a stable system formed by 
core and shell made of repulsive phantom energy. It means that a system 
constituted only by dark energy is able to achieve an equilibrium state. The 
second one reinforces this last conclusion since that it shows that even a 
black hole can be formed by a system formed exclusively by dark energy.

Thus, solving equation (\ref{M2}) for $\dot R(\tau)$ we can integrate 
$\dot R(\tau)$ and obtain $R(\tau)$, which are
shown in the figures \ref{fig7a} and \ref{fig8a}, for the cases
where the core and shell are made of dark energy.

\begin{table}
\caption{\label{tab:table3}This table summarizes all the configurations found 
in our analysis. The figures are shown only for the new systems which results
were never shown before in the literature.}
\begin{ruledtabular}
\begin{tabular}{ccccc}
Case & Interior Energy & Shell Energy & Figures & Structures \\
\hline
A & Standard           & Standard           &  & Black Hole \\
B & Standard           & Dark               &  & Black Hole \\
C & Standard           & Repulsive Phantom  &  & Black Hole/Dispersion \\
D & Dark               & Standard           & \ref{fig2},\ref{fig3},\ref{fig4},\ref{fig5} & Gravastar\\
E & Dark               & Dark               &  & Black Hole/Dispersion\\
F & Dark               & Repulsive Phantom  &  & Black Hole/Dispersion\\
G & Repulsive Phantom  & Standard           &  & Gravastar\\
H & Repulsive Phantom  & Dark               & \ref{fig8} & Black Hole\\
I & Repulsive Phantom  & Repulsive Phantom  & \ref{fig7} & Gravastar\\
J & Attractive Phantom & Standard           &  & None Structure \\
K & Attractive Phantom & Dark               &  & Black Hole \\
L & Attractive Phantom & Repulsive Phantom  &  & None Structure \\
\end{tabular}
\end{ruledtabular}
\end{table}

\section{Conclusions}

In this paper, we have studied the problem of the stability of gravastars by
constructing dynamical three-layer models  of VW \cite{VW04},
which consists of an internal anisotropic dark energy fluid, a dynamical infinitely thin  shell of
perfect fluid with the equation of state $p = (1-\gamma)\sigma$, and an external Schwarzschild space.

\begin{figure}
\vspace{.2in}
\centerline{\psfig{figure=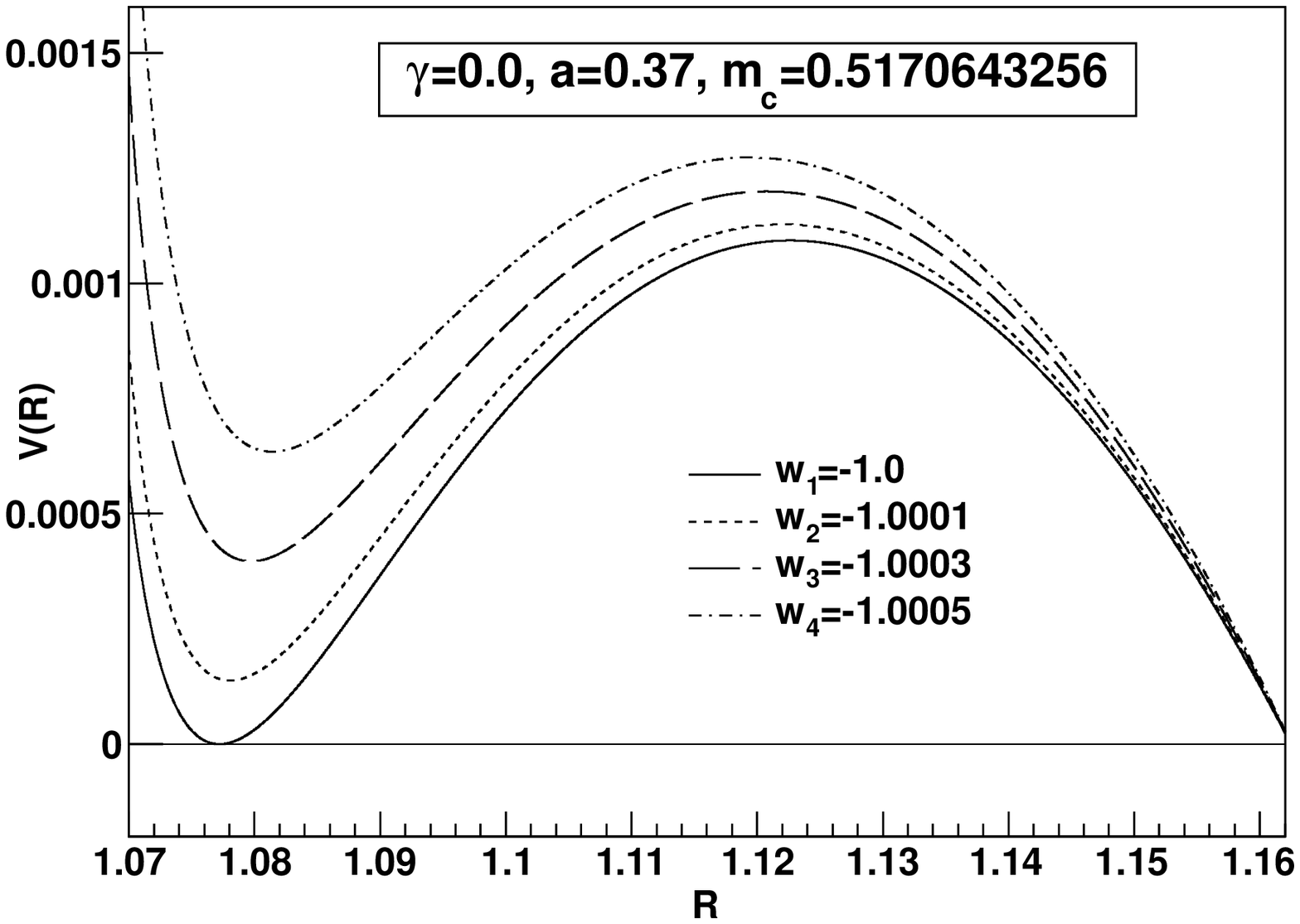,width=3.3truein,height=3.0truein}\hskip
.25in \psfig{figure=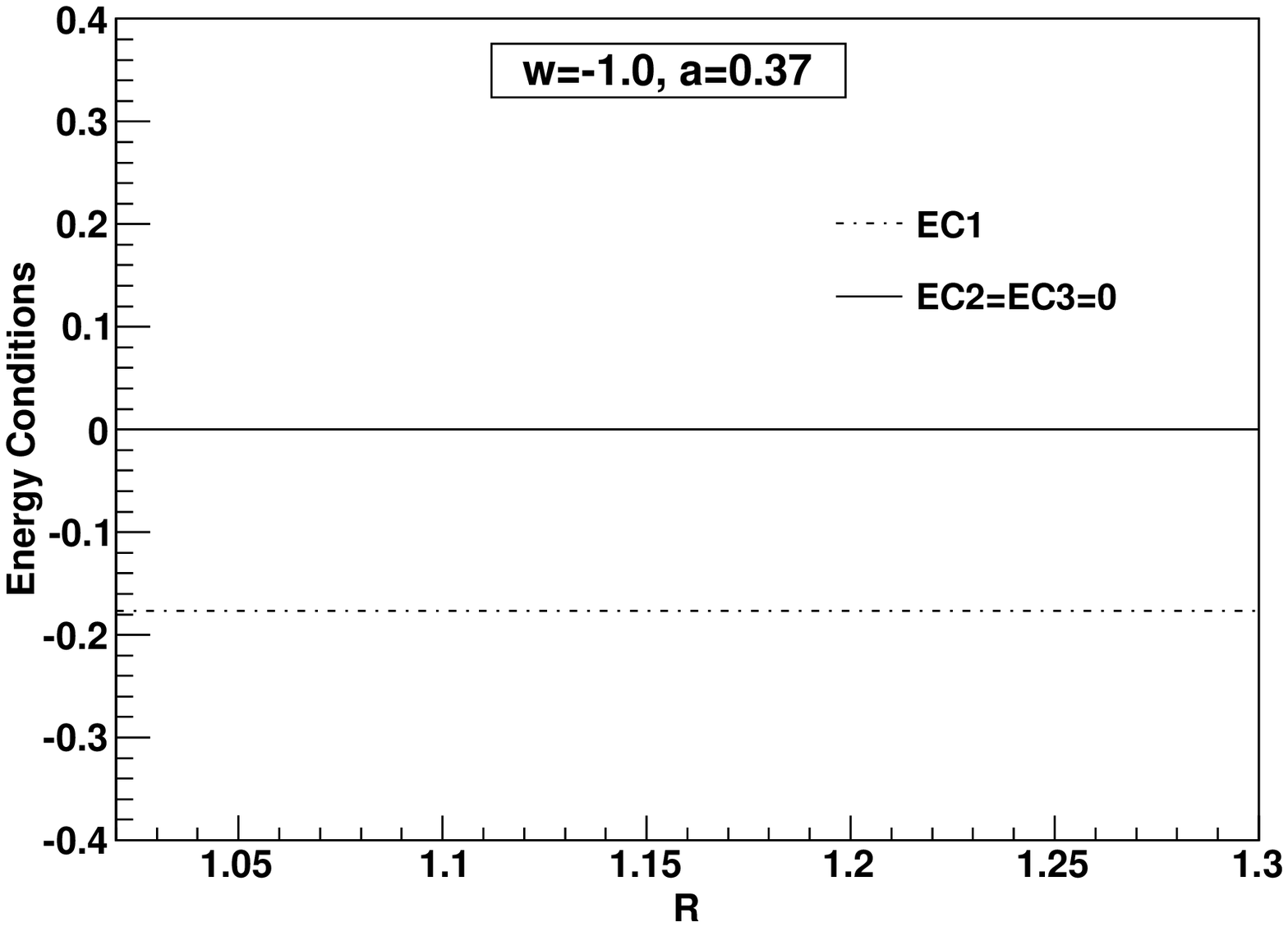,width=3.3truein,height=3.0truein}
\hskip .5in} \caption{The potential $V(R)$ and the energy conditions EC1$\equiv \rho+p_r+2p_t$, 
EC2$\equiv \rho+p_r$ and EC3$\equiv \rho+p_t$, for $\gamma=0$,
$\omega=-1,\; -1.0001,\; -1.0003, \; -1.0005$, $a=0.37$ and $m_c=0.5170643256$. {\bf Case D: $\omega=-1$}}
\label{fig2}
\end{figure}

\begin{figure}
\vspace{.2in}
\centerline{\psfig{figure=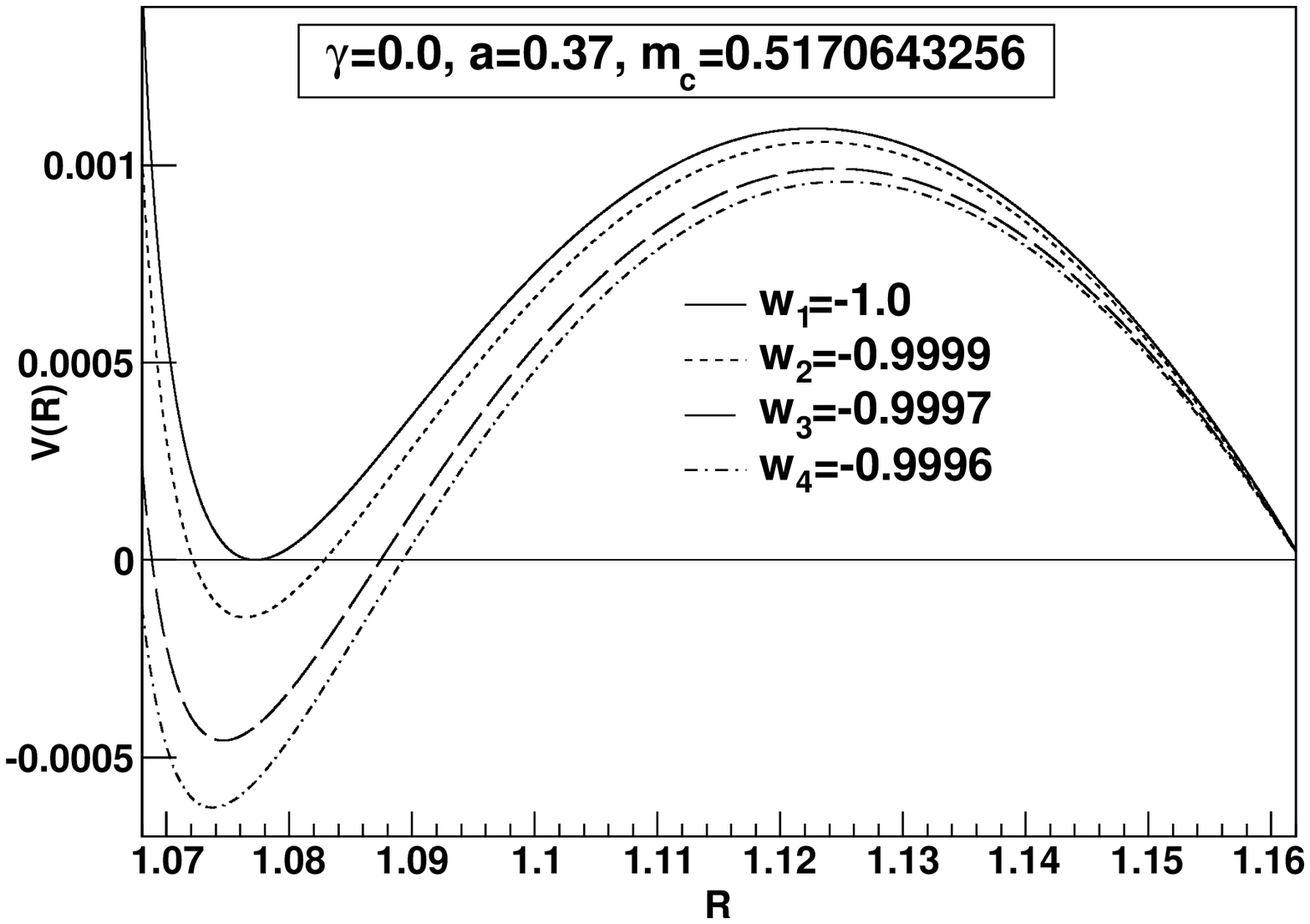,width=3.3truein,height=3.0truein}\hskip
.25in \psfig{figure=energycondwm1a0v37.eps,width=3.3truein,height=3.0truein}
\hskip .5in} \caption{The potential $V(R)$ and the energy conditions EC1$\equiv \rho+p_r+2p_t$, 
EC2$\equiv \rho+p_r$ and EC3$\equiv \rho+p_t$, for $\gamma=0$,
$\omega=-1,\; -0.9999,\; -0.9997, \; -0.9996$, $a=0.37$ and $m_c=0.5170643256$. {\bf Case D: $\omega=-1$}}
\label{fig3}
\end{figure}

\begin{figure}
\vspace{.2in}
\centerline{\psfig{figure=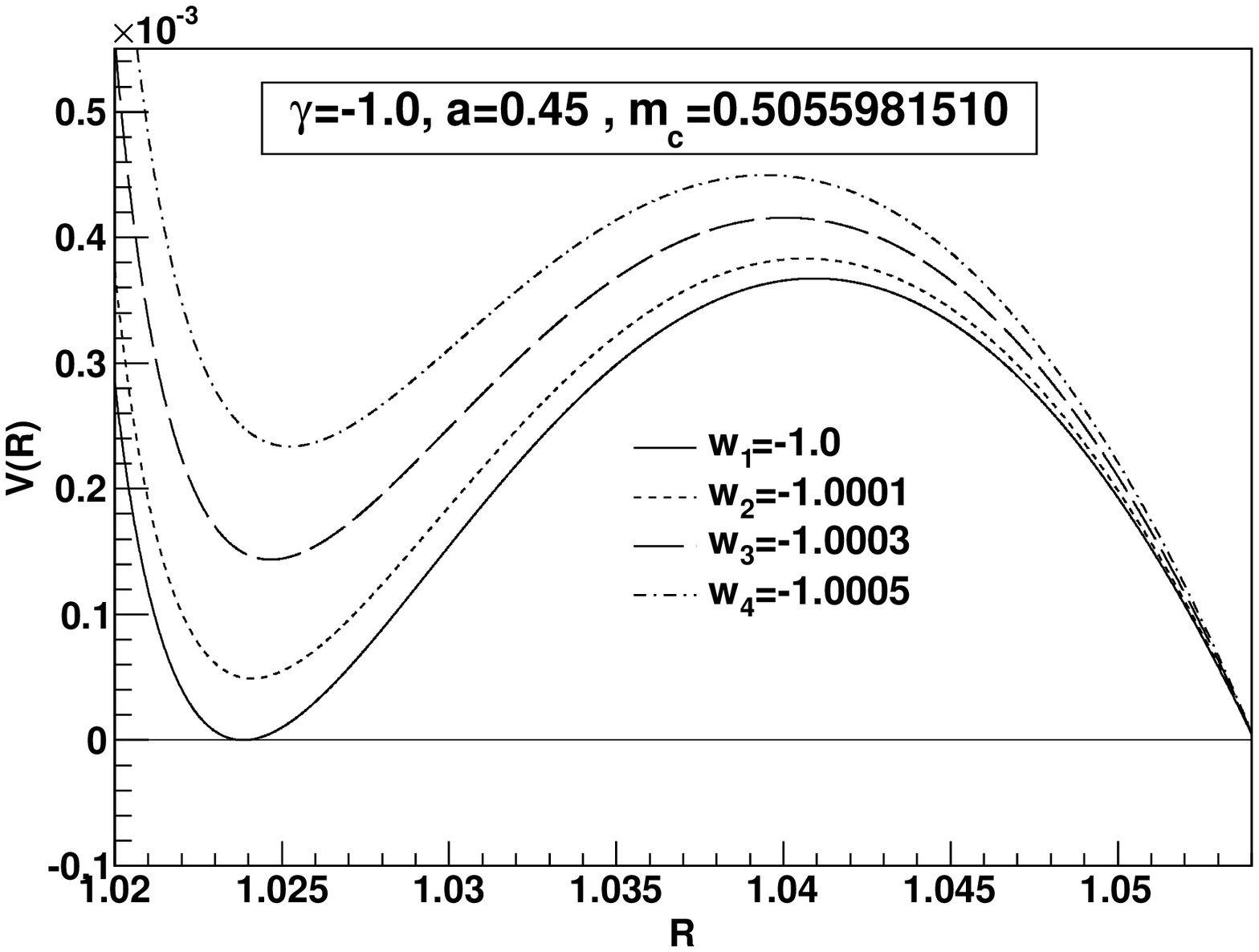,width=3.3truein,height=3.0truein}\hskip
.25in \psfig{figure=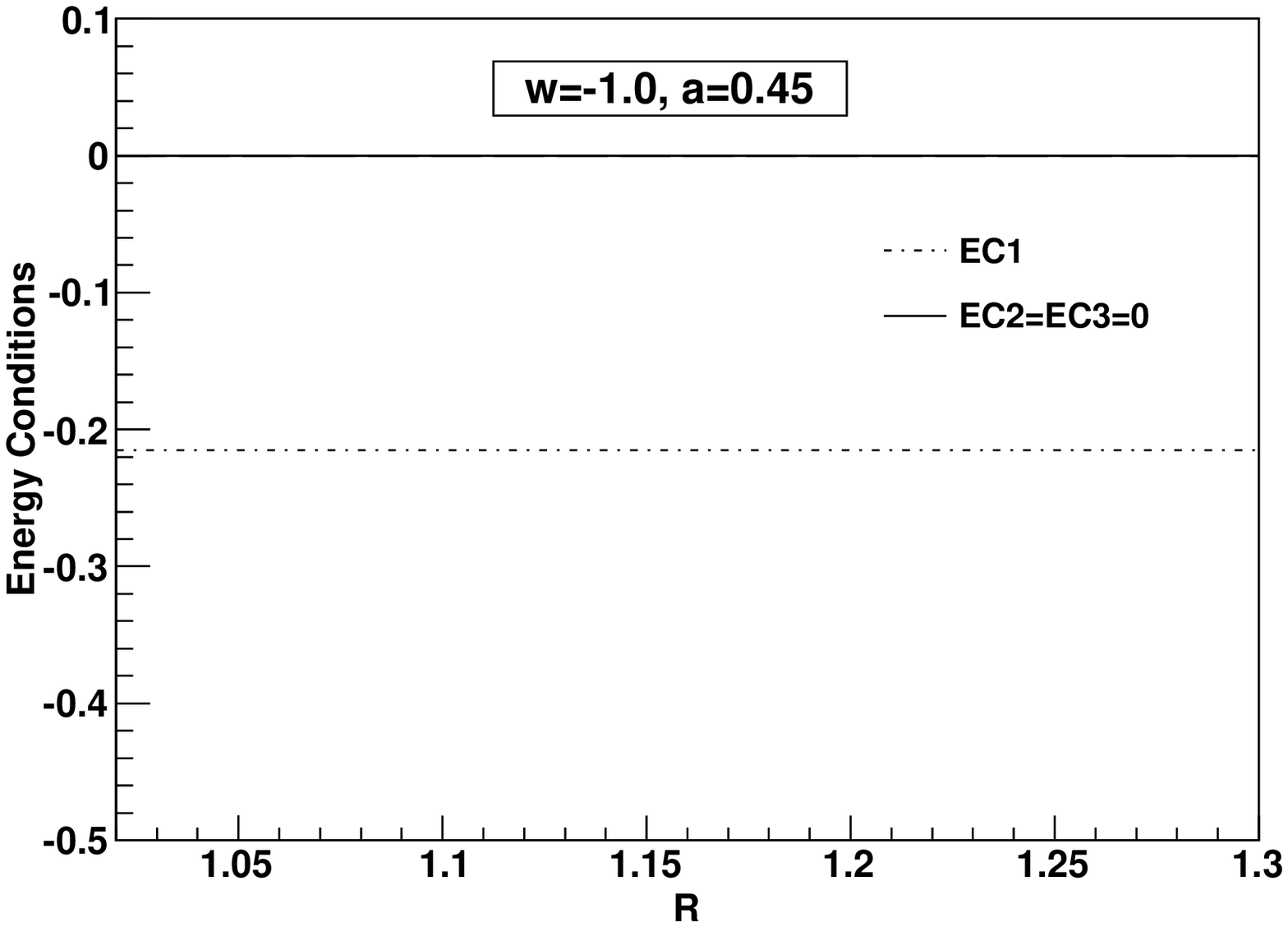,width=3.3truein,height=3.0truein}
\hskip .5in} \caption{The potential $V(R)$ and the energy conditions EC1$\equiv \rho+p_r+2p_t$, 
EC2$\equiv \rho+p_r$ and EC3$\equiv \rho+p_t$, for $\gamma=0$,
$\omega=-1,\; -1.0001,\; -1.0003, \; -1.0005$, $a=0.45$ and $m_c=0.5055981510$. {\bf Case D: $\omega=-1$}}
\label{fig4}
\end{figure}

\begin{figure}
\vspace{.2in}
\centerline{\psfig{figure=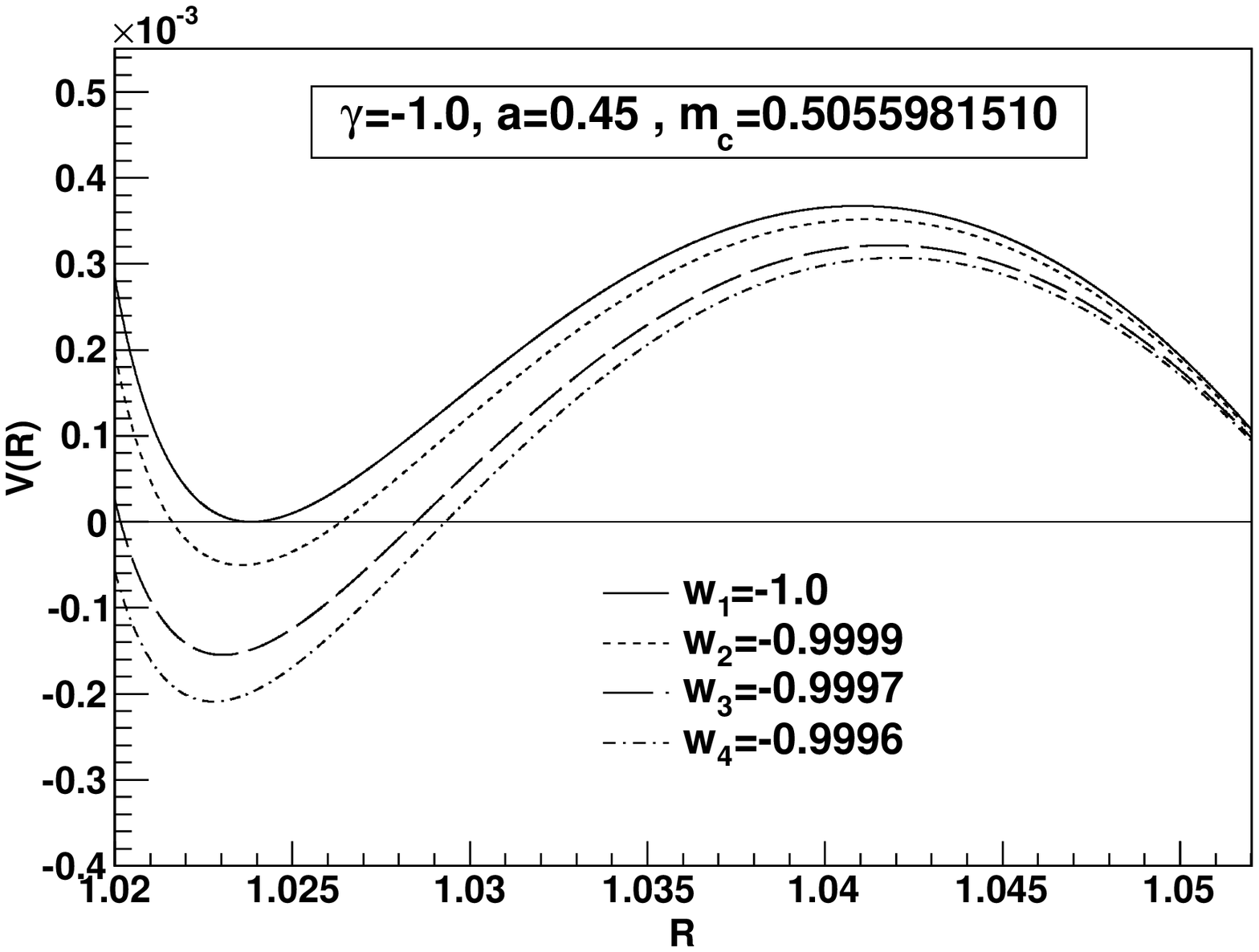,width=3.3truein,height=3.0truein}\hskip
.25in \psfig{figure=energycondwm1a0v45.eps,width=3.3truein,height=3.0truein}
\hskip .5in} \caption{The potential $V(R)$ and the energy conditions EC1$\equiv \rho+p_r+2p_t$, 
EC2$\equiv \rho+p_r$ and EC3$\equiv \rho+p_t$, for $\gamma=-1$,
$\omega=-1,\; -0.9999,\; -0.9997,\; -0.9996$, $a=0.45$ and $m_c=0.5055981510$. 
{\bf Case D: $\omega=-1$}}
\label{fig5}
\end{figure}

\begin{figure}
\vspace{.2in}
\centerline{\psfig{figure=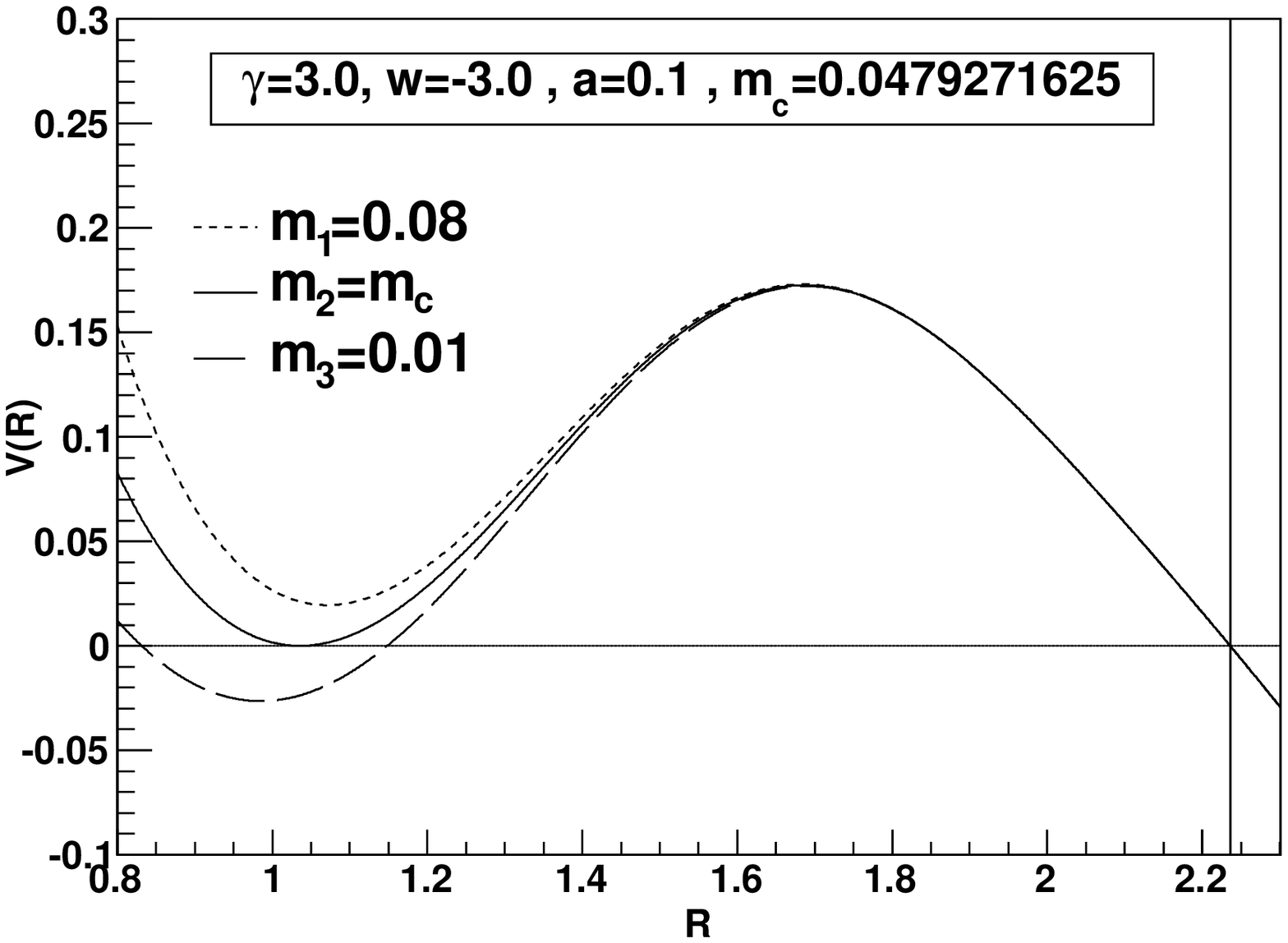,width=3.3truein,height=3.0truein}\hskip
.25in \psfig{figure=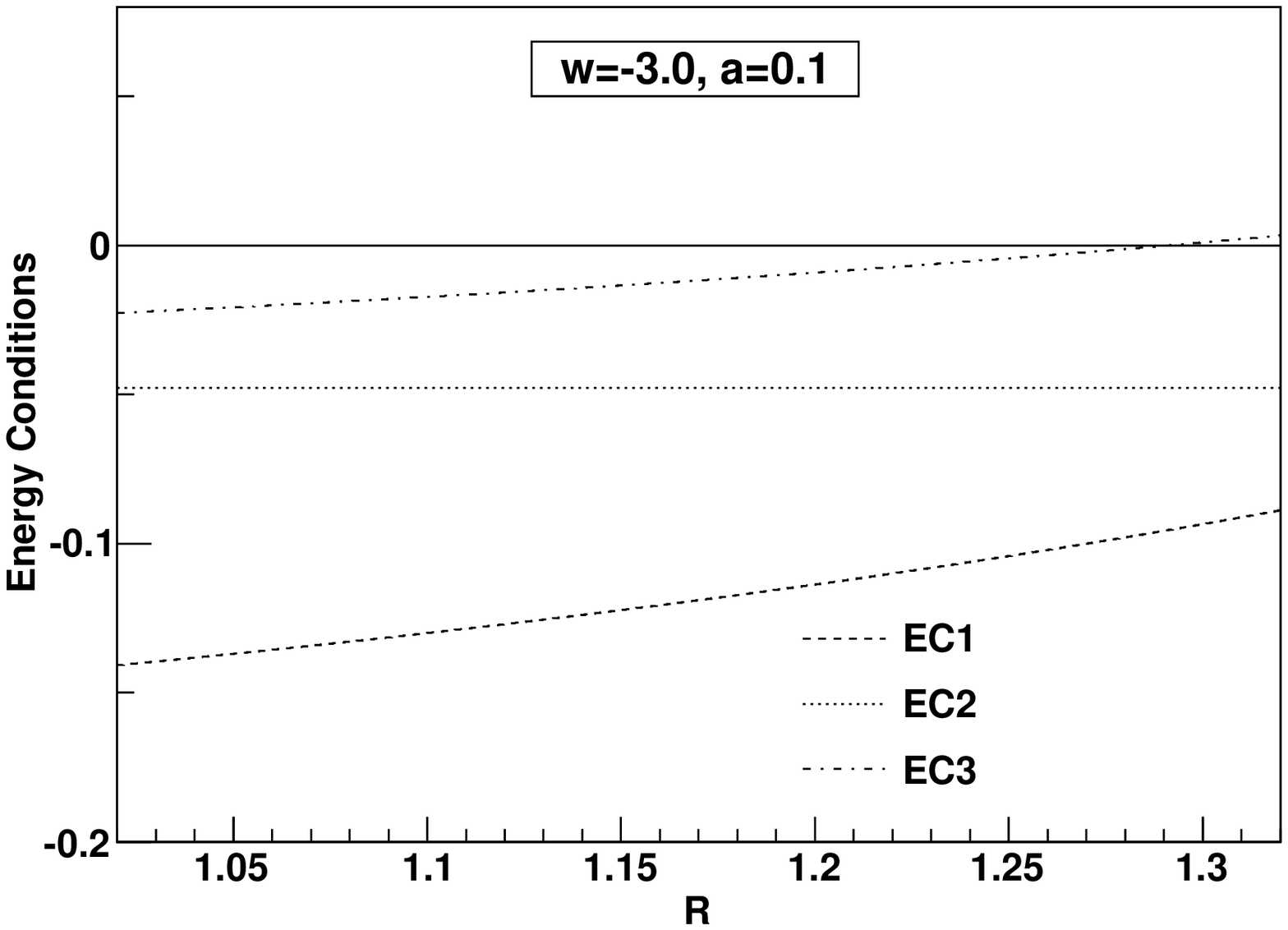,width=3.3truein,height=3.0truein}
\hskip .5in} \caption{The potential $V(R)$ and the energy conditions EC1$\equiv \rho+p_r+2p_t$, 
EC2$\equiv \rho+p_r$ and EC3$\equiv \rho+p_t$, for $\gamma=3$,
$\omega=-3$, $a=0.1$ and $m_c=0.0479271625$. The righter vertical straight line
denotes the radius of the horizon $R_h=1/\sqrt{2a}$. {\bf Case I}}
\label{fig7}
\end{figure}

\begin{figure}
\vspace{.2in}
\centerline{\psfig{figure=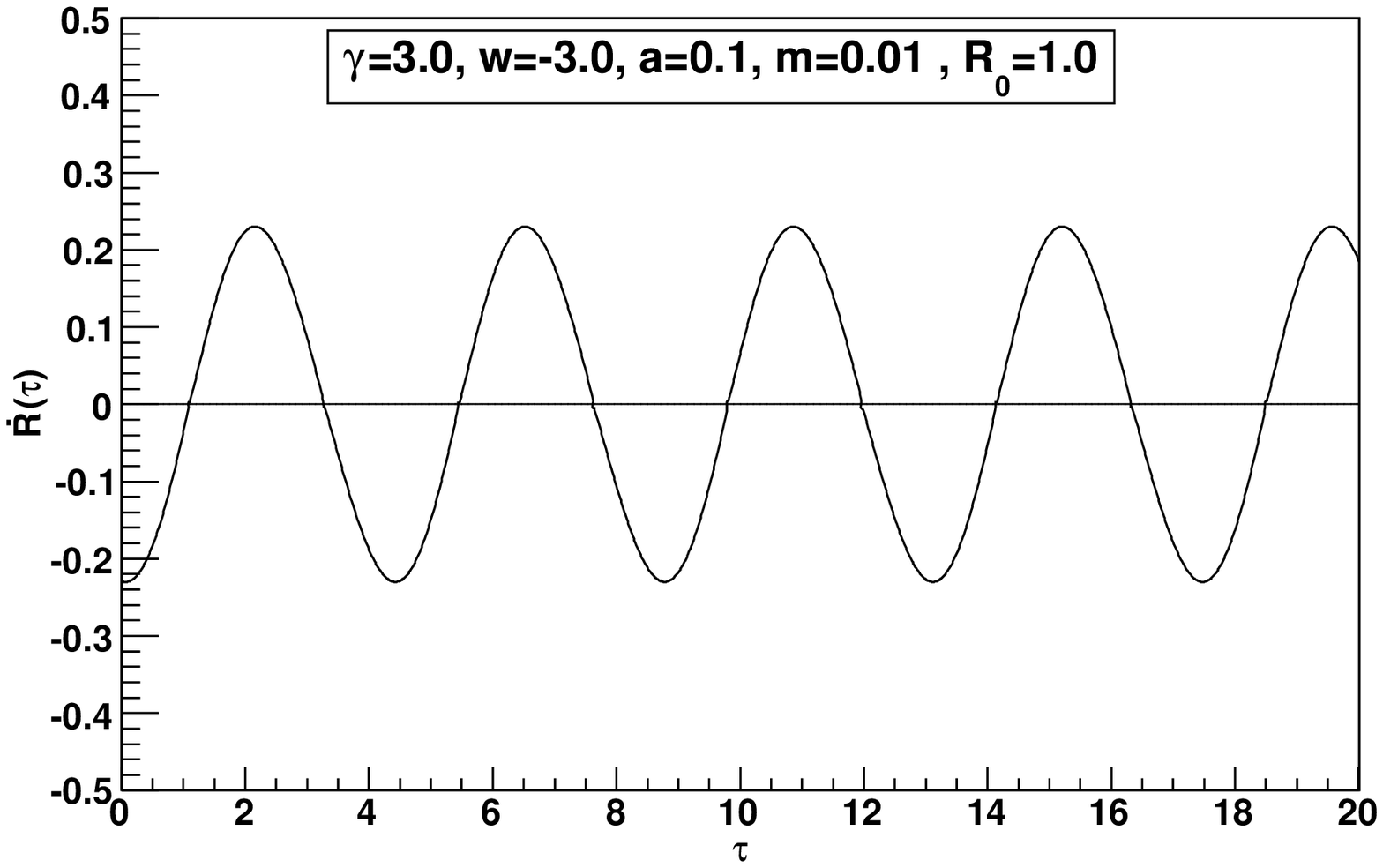,width=3.3truein,height=3.0truein}\hskip
.25in \psfig{figure=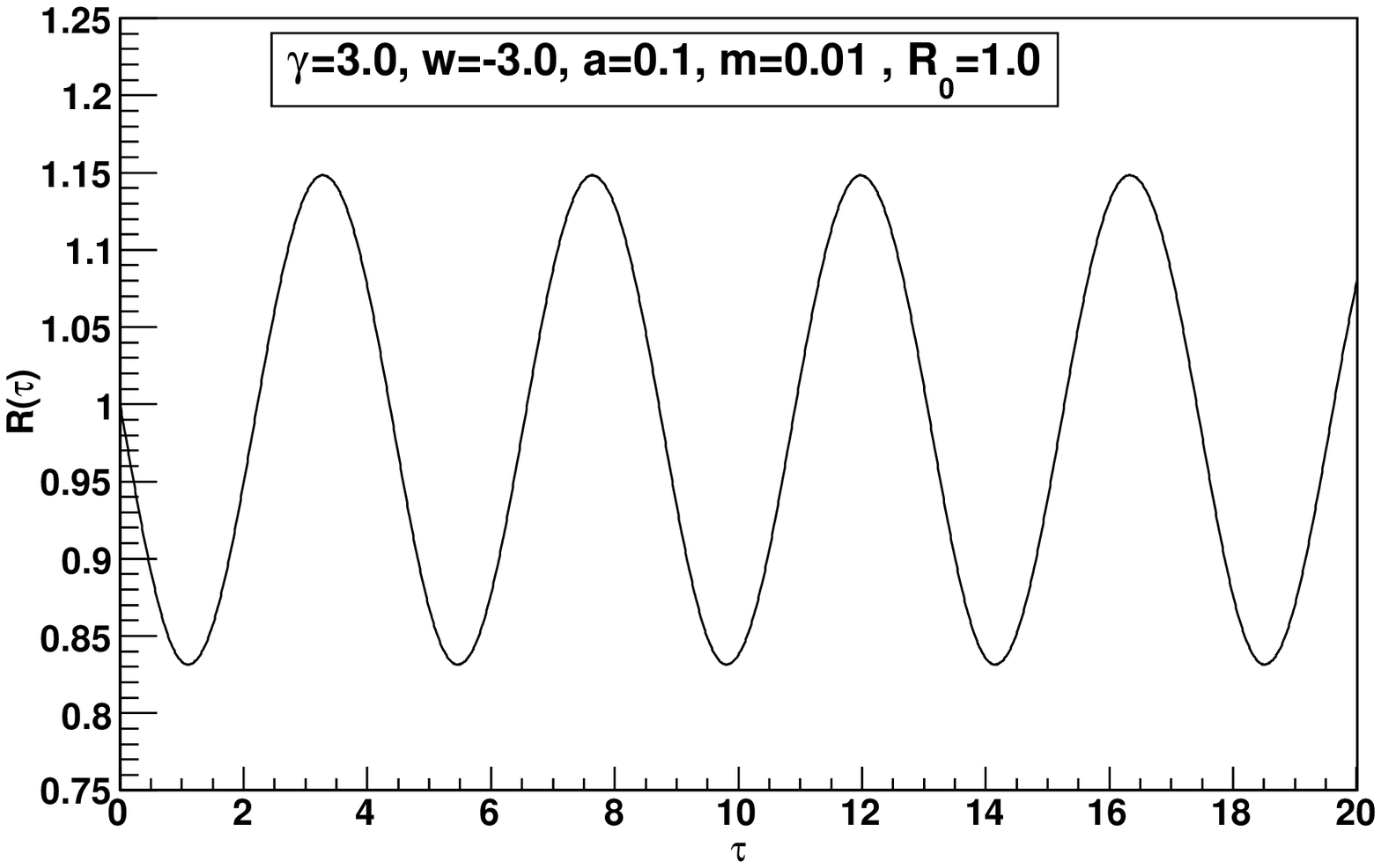,width=3.3truein,height=3.0truein}
\hskip .5in} \caption{These figures show the dynamical evolution of collapse
of a gravastar, forming a "bounded excursion" gravastar.
We have assumed the values $\gamma=3.0$, $\omega=-3.0$, $a=0.1$, $m=0.01$ and initial radius $R_0=R(\tau=0)=1.0$.}
\label{fig7a}
\end{figure}

\begin{figure}
\vspace{.2in}
\centerline{\psfig{figure=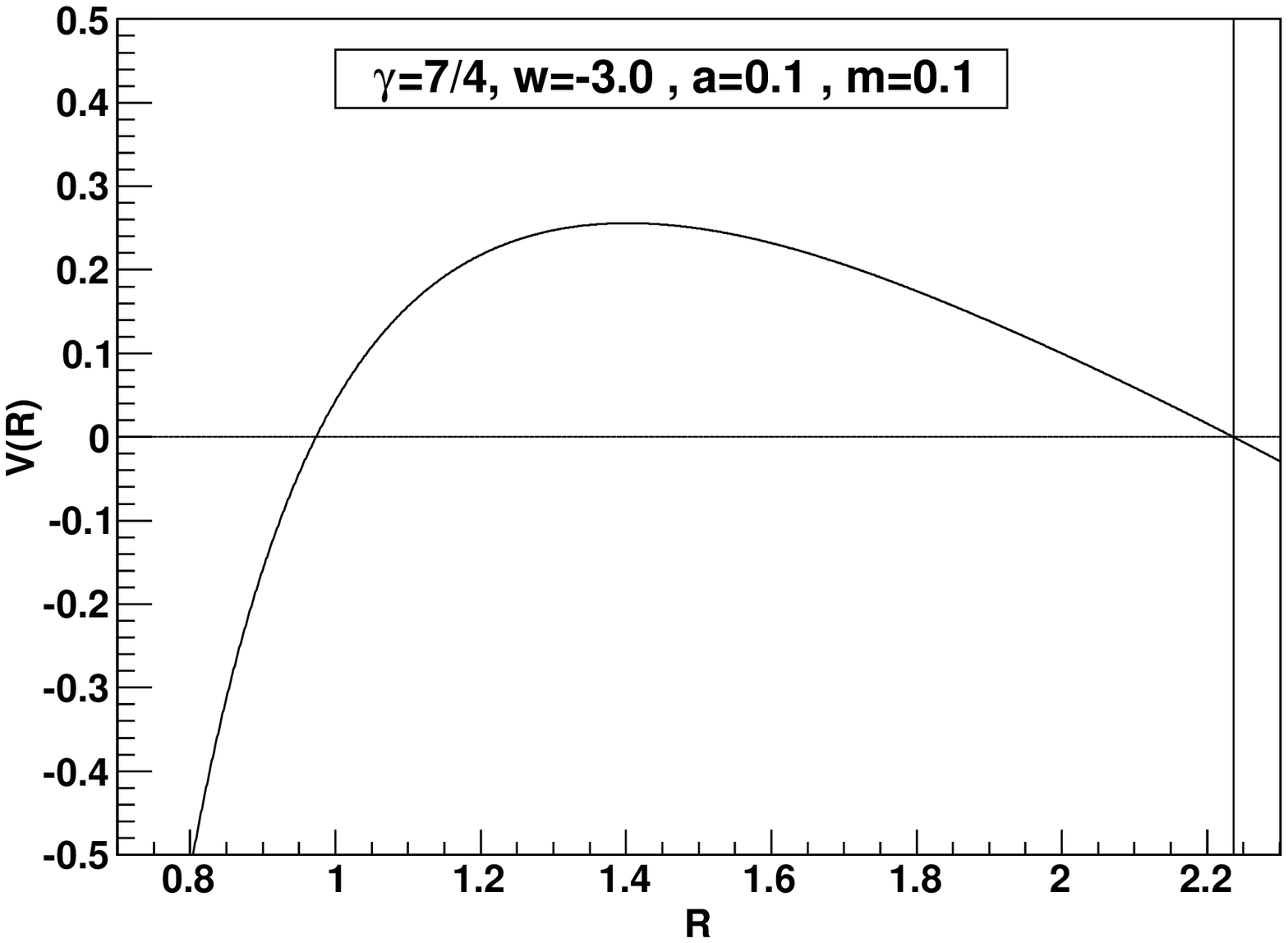,width=3.3truein,height=3.0truein}\hskip
.25in \psfig{figure=energycondwm3a0v1.eps,width=3.3truein,height=3.0truein}
\hskip .5in} \caption{The potential $V(R)$ and the energy conditions EC1$\equiv \rho+p_r+2p_t$, 
EC2$\equiv \rho+p_r$ and EC3$\equiv \rho+p_t$, for $\gamma=7/4$,
$\omega=-3$, $a=0.1$ and $m_c=0.1$. The righter vertical straight line
denotes the radius of the horizon $R_h=1/\sqrt{2a}$. {\bf Case H}}
\label{fig8}
\end{figure}

\begin{figure}
\vspace{.2in}
\centerline{\psfig{figure=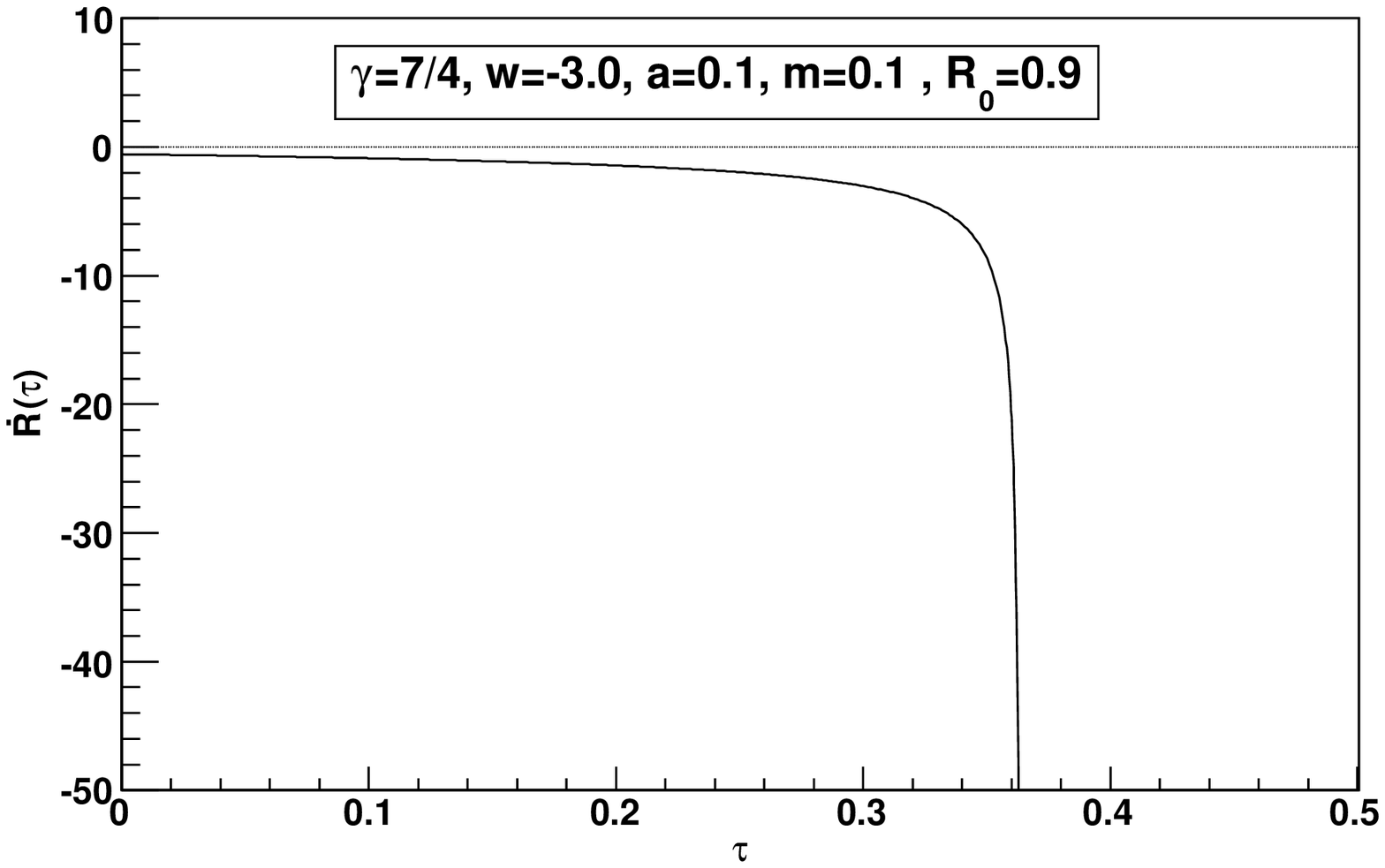,width=3.3truein,height=3.0truein}\hskip
.25in \psfig{figure=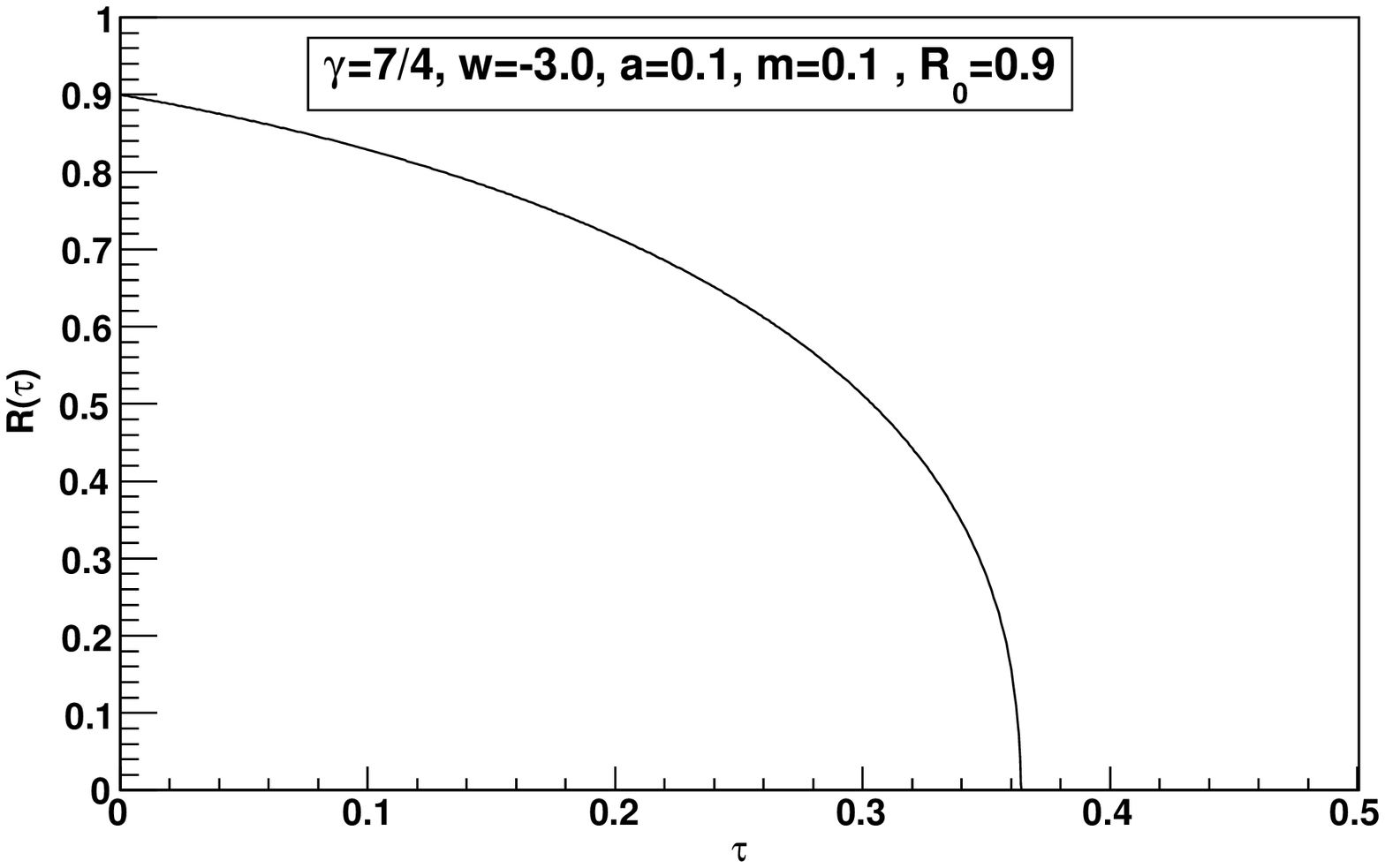,width=3.3truein,height=3.0truein}
\hskip .5in} \caption{These figures show the dynamical evolution of collapse
of a gravastar, forming a black hole.
We have assumed the values $\gamma=7/4$, $\omega=-3.0$, $a=0.1$, $m=0.1$ and initial radius $R_0=R(\tau=0)=0.9$.}
\label{fig8a}
\end{figure}

We also have shown that the isotropy or anisotropy
of the interior fluid may affect the gravastar formation. 
The formation is favored when the
tangential pressure is greater than the radial pressure, at least in the
neighborhood of the isotropic case ($\omega=-1$).
See figures \ref{fig2}, \ref{fig3}, \ref{fig4}, \ref{fig5},
where $\omega=-1$ represent the isotropic dark energy fluid and $\omega \ne -1$
represent the anisotropic dark energy fluid. 

We have shown explicitly that the final output can be a black
hole, a "bounded excursion" stable gravastar  or an anisotropic
dark energy 
spacetime, depending on the total mass $m$ of the system, the parameter 
$\omega$, 
the constant $a$, the parameter   $\gamma$ and
the initial position $R_{0}$ of the dynamical shell. 
All the results can be summarized in Table III. 
An interesting result that we have found is that we can have
gravastar and even black hole formation with an interior and thin shell 
dark energy.

\begin{acknowledgments}
The financial assistance from 
FAPERJ/UERJ (MFAdaS) are gratefully acknowledged. The
author (RC) acknowledges the financial support from FAPERJ (no.
E-26/171.754/2000, E-26/171.533/2002 and E-26/170.951/2006). 
The authors (RC and MFAdaS) also acknowledge the financial support from 
Conselho Nacional de Desenvolvimento Cient\'{\i}fico e Tecnol\'ogico - 
CNPq - Brazil.  The author (MFAdaS) also acknowledges the financial support
from Financiadora de Estudos e Projetos - FINEP - Brazil (Ref. 2399/03).
\end{acknowledgments}

\end{document}